\documentclass[a4paper,12pt]{article}
\usepackage[colorlinks=true,urlcolor=blue,linkcolor=red,citecolor=blue,anchorcolor=blue,breaklinks]{hyperref}
\usepackage{amsmath,amssymb,amsthm}
\usepackage[hyphenbreaks]{breakurl}
\usepackage[labelfont=bf,labelsep=period]{caption}
\usepackage{fancyhdr}
\begin{document}
\parindent 0cm

\begin{center}
\textbf{Tables of Quantiles of the Distribution of the Empirical Chiral Index\\ in the Case of the Uniform Law and in the Case of the Normal Law}
\end{center}
\vskip 0.5cm

\begin{center}
\textbf{Michel Petitjean}
\end{center}
\vskip 0.5cm

\small{$^1$ Universit\'e de Paris, BFA, CNRS UMR 8251, INSERM ERL U1133, F-75013 Paris, France.\\
       $^{2}$ E-p\^ole de G\'enoinformatique, CNRS UMR 7592, Institut Jacques Monod, 75013 Paris, France.}\\
       \textit{E-mail}: petitjean.chiral@gmail.com, michel.petitjean@univ-paris-diderot.fr.\\
       \href{http://petitjeanmichel.free.fr/itoweb.petitjean.html}{http://petitjeanmichel.free.fr/itoweb.petitjean.html}.\\
       \textit{ORCID}: \href{https://orcid.org/0000-0002-1745-5402}{0000-0002-1745-5402}.

\vskip 0.5cm
\textbf{Abstract}:
The empirical distribution of the chiral index is simulated for various sample sizes for the uniform law and and for the normal law.
The estimated quantiles $K_{0.90}$, $K_{0.95}$, $K_{0.98}$, and $K_{0.99}$, are tabulated for use in symmetry testing in the uniform case and in the normal case.



\section{Introduction}
\parindent 1cm

The chiral index $\chi$ is a quantitative measure of chirality which can be used as an asymmetry coefficient of multivariate distributions \cite{Petitjean2002,Petitjean2003}.
It takes values in the interval $\lbrack 0;1 \rbrack$.
The chiral index is null if and only if there is a mirror symmetry.
It is insensitive to isometries and scaling.
In the case of a $d$-variate distribution, its upper bound $\chi^{*}$ depends on the dimensionality $d$.
$\chi^{*}=1/2$ when $d=1$ \cite{Petitjean2002}, $\chi^{*} \in \lbrack 1-1/\pi;1-1/2\pi \rbrack$ when $d=2$ \cite{Coppersmith2005}, and $\chi^{*} \in \lbrack 1/2;1\rbrack$ when $d\ge 3$ \cite{Petitjean2008}.
As a consequence of a convergence theorem \cite{Petitjean2002}, the empirical chiral index of a mirror symmetric distribution converges to $0$, provided that the parent distribution has a finite and non null inertia.
So, the chiral index can be used for testing the symmetry of the parent law.
The purpose of this paper is to provide tables of the estimated quantiles $K_{0.90}$, $K_{0.95}$, $K_{0.98}$, and $K_{0.99}$, in the univariate uniform case and in the normal case.

\section{Theory}

The chiral index $\chi$ of an univariate distribution is given in equation \ref{CHI1} \cite{Petitjean2002,Petitjean2003}:
\begin{equation}
\label{CHI1}
\chi = ( 1 + r_m ) / 2
\end{equation}
where $r_m$ is the lower bound of the correlation coefficient between two identical distributions, taken over all their joint distributions.
$r_m$ cannot be positive, $\chi = 0$ if and only if the distribution is symmetric, and its upper bound is $\chi^{*}=1/2$.
This latter is asymptotically reached for the Bernouilli distribution with parameter tending to $0$ or $1$ \cite{Petitjean2002}.

\vskip 0.5cm
We consider a sample of size $n$ from a parent distribution with finite and non null inertia.
\\
$X_{i:n} \le ... \le X_{n:n}$ are the order statistics, $\bar{X}$ is their mean, and $\sigma$ is their standard deviation. 
$M_i=(X_{n+1-i:n}+X_{i:n})/2$ is the midrange of the $i^{th}$ interval, and
$L_i=(X_{n+1-i:n}-X_{i:n})/2$ is the half range length of the $i^{th}$ interval,
$i=1...n$. The $i^{th}$ interval and the $(n-i)^{th}$ interval are identical
except in their orientation. Thus $L_i$ is negative for $i>(n/2)$.
The chiral index can be expressed from the studentized variance of the midranges or of the half ranges lengths: see equations 2.9.4 and 2.9.5 in \cite{Petitjean2003}.
\begin{equation}
\chi = \Bigg[ \sum\limits_{i=1}^{i=n} M_i^{2} - n\cdot \bar{X}^{2} \Bigg] \Big/ (n\sigma^{2})
\end{equation}
\begin{equation}
\chi = 1 - \Bigg[ \sum\limits_{i=1}^{i=n} L_i^{2} \Bigg] \Big/ (n\sigma^{2})
\end{equation}

For practical computations, it is much simpler to use the expression of $r_m$ in equation \ref{CHI1}, for which the minimal correlation exists and is such that the sequence of observations sorted in increasing order is correlated with the sequence of observations sorted in decreasing order \cite{Petitjean2003,Petitjean1997}.
Thus, the algorithm to compute $\chi$ is very simple:

\parindent 1cm
\vskip 0.25cm
(a) Sort the sequence of $n$ observations in increasing order

(b) Correlate it with the sequence sorted in decreasing order

(c)  Add $1$ and then divide the result by $2$.
\vskip 0.25cm

\parindent 0cm
Step (a) is operating in $O(n\cdot Log(n))$ and step (b) in $O(n)$,
and the global algorithm is optimal worst-case in $O(n\cdot Log(n))$.
The whole computation is easy on a pocket calculator.
\parindent 1cm

\section{Simulations and results}

Table \ref{TAB1} was generated for the continuous uniform law $U(0,1)$.
Table \ref{TAB2} was generated for the normal law $N(0,1)$.
For each sample size $n$ from $U(0,1)$ or from $N(0,1)$, we compute one chiral index, i.e., one observation from the distribution of the chiral index.
So, to approach this distribution, 10000 observations of the chiral index are generated.
For each set of 10000 observations of the chiral index, four quantiles are computed as follow ($\chi_{i:10000}$ denotes the $i^{th}$ order satistic):
\begin{eqnarray}
K_{0.90} = ( \chi_{9000:10000} + \chi_{9001:10000} ) / 2 \\
K_{0.95} = ( \chi_{9500:10000} + \chi_{9501:10000} ) / 2 \\
K_{0.98} = ( \chi_{9800:10000} + \chi_{9801:10000} ) / 2 \\
K_{0.99} = ( \chi_{9900:10000} + \chi_{9901:10000} ) / 2
\end{eqnarray}
The whole simulation is repeated 100 times, and both the mean values
$\bar{K}_{0.90}$, $\bar{K}_{0.95}$, $\bar{K}_{0.98}$, and $\bar{K}_{0.99}$,
and the estimated standard deviations
$S_{K_{0.90}}$, $S_{K_{0.95}}$, $S_{K_{0.98}}$, and $S_{K_{0.99}}$
of the four distributions of the quantiles are calculated and reported in tables \ref{TAB1} and \ref{TAB2}.

\vskip 0.5cm
The random generator is a long period pseudo-random congruential multiplicative one, used in the routine \textit{g05saf} of the NAG library \cite{NAG2017}.
All data of both tables were generated sequentially within a single executable program, without reinitializing the random generator.

\vskip 0.5cm
Tables \ref{TAB1} or \ref{TAB2} can be used as follows.
Compute the chiral index $\chi$ of your sample of size $n$.
\\
$\bullet$ If you assume that your sample is issued from the continuous uniform law $U(a,b)$, $a<b$, look in Table \ref{TAB1} if $\chi$ exceeds the quantile value at the desired confidence level (e.g. $0.95$).
If yes, you may reject the symmetry assumption, and thus you may reject the uniform law  assumption.
\\
$\bullet$ If you assume that your sample is issued from the normal law $N(m,s)$, look in Table \ref{TAB2} if $\chi$ exceeds the quantile value at the desired confidence level (e.g. $0.95$).
If yes, you may reject the symmetry assumption, and thus you may reject the normal law  assumption.

\newpage
\parindent 0cm

\begin{table}[!ht]
\begin{center}
\caption{Simulations for the uniform law $U(0,1)$.
         $n$: sample size.\\
         $\bar{K}_{0.90}$, $\bar{K}_{0.95}$, $\bar{K}_{0.98}$, $\bar{K}_{0.99}$: mean values of estimated quantiles over 100 experiments.\\
         $S_{K_{0.90}}$, $S_{K_{0.95}}$, $S_{K_{0.98}}$, $S_{K_{0.99}}$: estimated standard deviations of the quantiles distributions.}
\label{TAB1}
\end{center}
\end{table}
\vspace*{-0.75cm}
\begin{tabular}{|r|r|r|r|r|r|r|r|r|}
\hline
 $n$ & $\bar{K}_{0.90}$ & $\bar{K}_{0.95}$ & $\bar{K}_{0.98}$ & $\bar{K}_{0.99}$
     & $S_{K_{0.90}}$ & $S_{K_{0.95}}$ & $S_{K_{0.98}}$ & $S_{K_{0.99}}$ \\
\hline
       3 & 0.212764 & 0.231432 & 0.242661 & 0.246342 & 0.001020 & 0.000774 & 0.000583 & 0.000320 \\
       4 & 0.185211 & 0.229221 & 0.268552 & 0.287850 & 0.002169 & 0.002272 & 0.002310 & 0.002153 \\
       5 & 0.154894 & 0.195716 & 0.242379 & 0.270442 & 0.001717 & 0.002531 & 0.003358 & 0.003770 \\
       6 & 0.138459 & 0.174426 & 0.214915 & 0.242201 & 0.001695 & 0.001989 & 0.002476 & 0.003498 \\
       7 & 0.122632 & 0.156176 & 0.195287 & 0.221293 & 0.001495 & 0.001935 & 0.002647 & 0.003295 \\
       8 & 0.109606 & 0.139940 & 0.176472 & 0.201666 & 0.001339 & 0.001867 & 0.002653 & 0.003400 \\
       9 & 0.099866 & 0.127905 & 0.161753 & 0.185403 & 0.001367 & 0.001822 & 0.002665 & 0.003770 \\
      10 & 0.090742 & 0.116454 & 0.148675 & 0.171121 & 0.001184 & 0.001687 & 0.002358 & 0.003258 \\
      11 & 0.083883 & 0.107905 & 0.137897 & 0.158657 & 0.001161 & 0.001501 & 0.002188 & 0.003212 \\
      12 & 0.077375 & 0.099899 & 0.128247 & 0.148157 & 0.000985 & 0.001459 & 0.002142 & 0.002723 \\
      13 & 0.072242 & 0.093095 & 0.119908 & 0.138950 & 0.000913 & 0.001188 & 0.002112 & 0.002306 \\
      14 & 0.067347 & 0.087017 & 0.112577 & 0.130622 & 0.000790 & 0.001316 & 0.001573 & 0.002496 \\
      15 & 0.063401 & 0.081884 & 0.105740 & 0.123299 & 0.000871 & 0.001163 & 0.001712 & 0.002299 \\
      16 & 0.059669 & 0.077378 & 0.100094 & 0.116943 & 0.000788 & 0.001031 & 0.001639 & 0.002331 \\
      17 & 0.056655 & 0.073521 & 0.095419 & 0.111137 & 0.000743 & 0.001169 & 0.001649 & 0.002057 \\
      18 & 0.053513 & 0.069685 & 0.090491 & 0.105711 & 0.000691 & 0.001029 & 0.001668 & 0.002266 \\
      19 & 0.050912 & 0.066281 & 0.086226 & 0.100894 & 0.000624 & 0.000967 & 0.001355 & 0.002063 \\
      20 & 0.048617 & 0.063213 & 0.082350 & 0.096221 & 0.000623 & 0.000875 & 0.001360 & 0.001722 \\
      21 & 0.046461 & 0.060632 & 0.078984 & 0.092450 & 0.000607 & 0.000980 & 0.001557 & 0.001941 \\
      22 & 0.044460 & 0.058067 & 0.075878 & 0.088865 & 0.000586 & 0.000829 & 0.001383 & 0.001741 \\
      23 & 0.042783 & 0.055897 & 0.073019 & 0.085646 & 0.000673 & 0.000865 & 0.001218 & 0.001951 \\
      24 & 0.040955 & 0.053483 & 0.070111 & 0.082124 & 0.000591 & 0.000734 & 0.001247 & 0.001733 \\
      25 & 0.039345 & 0.051406 & 0.067428 & 0.079209 & 0.000463 & 0.000868 & 0.001282 & 0.001677 \\
      26 & 0.037959 & 0.049700 & 0.065121 & 0.076492 & 0.000457 & 0.000749 & 0.001253 & 0.001640 \\
      27 & 0.036764 & 0.048079 & 0.063067 & 0.074118 & 0.000524 & 0.000689 & 0.001010 & 0.001453 \\
      28 & 0.035401 & 0.046299 & 0.060849 & 0.071566 & 0.000507 & 0.000681 & 0.001274 & 0.001703 \\
      29 & 0.034266 & 0.044886 & 0.058923 & 0.069457 & 0.000519 & 0.000673 & 0.001147 & 0.001431 \\
      30 & 0.033099 & 0.043416 & 0.057067 & 0.067353 & 0.000409 & 0.000569 & 0.000986 & 0.001319 \\
      31 & 0.032121 & 0.042051 & 0.055308 & 0.065082 & 0.000429 & 0.000658 & 0.000945 & 0.001362 \\
      32 & 0.031084 & 0.040791 & 0.053830 & 0.063512 & 0.000405 & 0.000621 & 0.001171 & 0.001500 \\
      33 & 0.030233 & 0.039744 & 0.052259 & 0.061845 & 0.000389 & 0.000501 & 0.000990 & 0.001308 \\
      34 & 0.029371 & 0.038641 & 0.050966 & 0.060222 & 0.000427 & 0.000659 & 0.001019 & 0.001513 \\
      35 & 0.028665 & 0.037600 & 0.049672 & 0.058667 & 0.000424 & 0.000611 & 0.000893 & 0.001281 \\
      36 & 0.027763 & 0.036530 & 0.048173 & 0.057204 & 0.000354 & 0.000584 & 0.000949 & 0.001177 \\
      37 & 0.027185 & 0.035736 & 0.047072 & 0.055737 & 0.000371 & 0.000599 & 0.000971 & 0.001313 \\
      38 & 0.026346 & 0.034660 & 0.045839 & 0.054346 & 0.000359 & 0.000478 & 0.000817 & 0.001306 \\
      39 & 0.025722 & 0.033817 & 0.044709 & 0.053001 & 0.000307 & 0.000518 & 0.000888 & 0.001334 \\
      40 & 0.025175 & 0.033091 & 0.043765 & 0.051885 & 0.000352 & 0.000512 & 0.000833 & 0.001107 \\
      41 & 0.024560 & 0.032330 & 0.042778 & 0.050509 & 0.000328 & 0.000461 & 0.000848 & 0.001111 \\
      42 & 0.024013 & 0.031578 & 0.041900 & 0.049560 & 0.000275 & 0.000440 & 0.000796 & 0.001129 \\
      43 & 0.023479 & 0.030914 & 0.040873 & 0.048470 & 0.000314 & 0.000473 & 0.000764 & 0.001153 \\
      44 & 0.022904 & 0.030150 & 0.039905 & 0.047332 & 0.000296 & 0.000401 & 0.000707 & 0.000884 \\
      45 & 0.022436 & 0.029536 & 0.039130 & 0.046380 & 0.000308 & 0.000474 & 0.000796 & 0.001086 \\
\hline
\end{tabular}

\newpage
\textbf{Table 1.} Continued.
\vskip 0.5cm

\begin{tabular}{|r|r|r|r|r|r|r|r|r|}
\hline
      46 & 0.021912 & 0.028844 & 0.038205 & 0.045312 & 0.000271 & 0.000395 & 0.000677 & 0.001032 \\
      47 & 0.021513 & 0.028343 & 0.037557 & 0.044496 & 0.000293 & 0.000427 & 0.000736 & 0.000993 \\
      48 & 0.021034 & 0.027734 & 0.036730 & 0.043550 & 0.000261 & 0.000407 & 0.000606 & 0.000907 \\
      49 & 0.020630 & 0.027197 & 0.035990 & 0.042803 & 0.000279 & 0.000415 & 0.000791 & 0.001129 \\
      50 & 0.020233 & 0.026654 & 0.035312 & 0.041995 & 0.000260 & 0.000389 & 0.000583 & 0.000858 \\
      51 & 0.019919 & 0.026211 & 0.034795 & 0.041392 & 0.000283 & 0.000397 & 0.000710 & 0.000930 \\
      52 & 0.019437 & 0.025630 & 0.034018 & 0.040360 & 0.000289 & 0.000382 & 0.000645 & 0.000909 \\
      53 & 0.019145 & 0.025207 & 0.033497 & 0.039760 & 0.000257 & 0.000393 & 0.000566 & 0.000738 \\
      54 & 0.018779 & 0.024782 & 0.032827 & 0.038877 & 0.000271 & 0.000368 & 0.000610 & 0.000911 \\
      55 & 0.018481 & 0.024346 & 0.032247 & 0.038246 & 0.000234 & 0.000399 & 0.000669 & 0.000914 \\
      56 & 0.018195 & 0.023973 & 0.031819 & 0.037913 & 0.000223 & 0.000348 & 0.000622 & 0.000879 \\
      57 & 0.017848 & 0.023510 & 0.031156 & 0.036994 & 0.000235 & 0.000317 & 0.000510 & 0.000796 \\
      58 & 0.017540 & 0.023125 & 0.030707 & 0.036498 & 0.000227 & 0.000330 & 0.000580 & 0.000791 \\
      59 & 0.017256 & 0.022719 & 0.030166 & 0.035924 & 0.000244 & 0.000360 & 0.000608 & 0.000911 \\
      60 & 0.016933 & 0.022354 & 0.029726 & 0.035475 & 0.000199 & 0.000327 & 0.000583 & 0.000872 \\
      61 & 0.016717 & 0.022031 & 0.029217 & 0.034741 & 0.000206 & 0.000330 & 0.000589 & 0.000812 \\
      62 & 0.016420 & 0.021642 & 0.028772 & 0.034233 & 0.000215 & 0.000315 & 0.000526 & 0.000739 \\
      63 & 0.016188 & 0.021349 & 0.028339 & 0.033746 & 0.000210 & 0.000340 & 0.000574 & 0.000838 \\
      64 & 0.015933 & 0.021102 & 0.028073 & 0.033321 & 0.000208 & 0.000322 & 0.000554 & 0.000765 \\
      65 & 0.015717 & 0.020733 & 0.027559 & 0.032849 & 0.000210 & 0.000341 & 0.000517 & 0.000701 \\
      66 & 0.015451 & 0.020362 & 0.027101 & 0.032306 & 0.000208 & 0.000292 & 0.000490 & 0.000679 \\
      67 & 0.015220 & 0.020104 & 0.026754 & 0.031902 & 0.000212 & 0.000319 & 0.000473 & 0.000708 \\
      68 & 0.014980 & 0.019793 & 0.026426 & 0.031379 & 0.000220 & 0.000301 & 0.000510 & 0.000687 \\
      69 & 0.014789 & 0.019509 & 0.026005 & 0.030905 & 0.000200 & 0.000311 & 0.000505 & 0.000675 \\
      70 & 0.014596 & 0.019222 & 0.025610 & 0.030422 & 0.000180 & 0.000279 & 0.000464 & 0.000701 \\
      71 & 0.014384 & 0.018999 & 0.025335 & 0.030181 & 0.000197 & 0.000306 & 0.000495 & 0.000731 \\
      72 & 0.014147 & 0.018706 & 0.024913 & 0.029688 & 0.000171 & 0.000258 & 0.000490 & 0.000679 \\
      73 & 0.013969 & 0.018462 & 0.024527 & 0.029260 & 0.000163 & 0.000254 & 0.000424 & 0.000756 \\
      74 & 0.013788 & 0.018167 & 0.024190 & 0.028830 & 0.000201 & 0.000281 & 0.000457 & 0.000682 \\
      75 & 0.013624 & 0.018002 & 0.023958 & 0.028546 & 0.000193 & 0.000268 & 0.000456 & 0.000623 \\
      76 & 0.013447 & 0.017747 & 0.023662 & 0.028168 & 0.000190 & 0.000273 & 0.000414 & 0.000566 \\
      77 & 0.013302 & 0.017557 & 0.023380 & 0.027778 & 0.000177 & 0.000261 & 0.000425 & 0.000659 \\
      78 & 0.013119 & 0.017334 & 0.023157 & 0.027641 & 0.000177 & 0.000285 & 0.000470 & 0.000709 \\
      79 & 0.012959 & 0.017119 & 0.022789 & 0.027187 & 0.000178 & 0.000272 & 0.000400 & 0.000622 \\
      80 & 0.012777 & 0.016868 & 0.022473 & 0.026816 & 0.000175 & 0.000272 & 0.000441 & 0.000686 \\
      81 & 0.012627 & 0.016715 & 0.022264 & 0.026520 & 0.000164 & 0.000264 & 0.000417 & 0.000662 \\
      82 & 0.012499 & 0.016512 & 0.022052 & 0.026281 & 0.000178 & 0.000289 & 0.000403 & 0.000529 \\
      83 & 0.012352 & 0.016319 & 0.021718 & 0.025897 & 0.000158 & 0.000248 & 0.000408 & 0.000606 \\
      84 & 0.012192 & 0.016114 & 0.021558 & 0.025664 & 0.000182 & 0.000264 & 0.000398 & 0.000553 \\
      85 & 0.012077 & 0.015989 & 0.021302 & 0.025331 & 0.000169 & 0.000262 & 0.000411 & 0.000573 \\
      86 & 0.011906 & 0.015734 & 0.020964 & 0.024974 & 0.000153 & 0.000249 & 0.000426 & 0.000618 \\
      87 & 0.011756 & 0.015554 & 0.020740 & 0.024612 & 0.000149 & 0.000214 & 0.000363 & 0.000521 \\
      88 & 0.011650 & 0.015398 & 0.020566 & 0.024557 & 0.000172 & 0.000220 & 0.000373 & 0.000492 \\
      89 & 0.011506 & 0.015238 & 0.020295 & 0.024188 & 0.000179 & 0.000242 & 0.000417 & 0.000572 \\
      90 & 0.011421 & 0.015099 & 0.020147 & 0.024046 & 0.000166 & 0.000230 & 0.000428 & 0.000545 \\
\hline
\end{tabular}

\newpage
\textbf{Table 1.} Continued.
\vskip 0.5cm

\begin{tabular}{|r|r|r|r|r|r|r|r|r|}
\hline
      91 & 0.011254 & 0.014856 & 0.019800 & 0.023658 & 0.000148 & 0.000235 & 0.000372 & 0.000564 \\
      92 & 0.011142 & 0.014747 & 0.019618 & 0.023369 & 0.000137 & 0.000223 & 0.000367 & 0.000542 \\
      93 & 0.011026 & 0.014585 & 0.019455 & 0.023258 & 0.000137 & 0.000218 & 0.000377 & 0.000570 \\
      94 & 0.010927 & 0.014433 & 0.019267 & 0.023014 & 0.000173 & 0.000265 & 0.000363 & 0.000512 \\
      95 & 0.010822 & 0.014315 & 0.019090 & 0.022762 & 0.000153 & 0.000213 & 0.000364 & 0.000586 \\
      96 & 0.010670 & 0.014148 & 0.018876 & 0.022528 & 0.000146 & 0.000221 & 0.000355 & 0.000482 \\
      97 & 0.010588 & 0.014029 & 0.018659 & 0.022190 & 0.000136 & 0.000200 & 0.000335 & 0.000470 \\
      98 & 0.010484 & 0.013846 & 0.018503 & 0.022106 & 0.000146 & 0.000224 & 0.000319 & 0.000460 \\
      99 & 0.010364 & 0.013718 & 0.018361 & 0.021858 & 0.000140 & 0.000224 & 0.000321 & 0.000513 \\
     100 & 0.010268 & 0.013574 & 0.018043 & 0.021550 & 0.000143 & 0.000230 & 0.000368 & 0.000502 \\
\hline
     110 & 0.009351 & 0.012355 & 0.016481 & 0.019638 & 0.000132 & 0.000204 & 0.000356 & 0.000485 \\
     120 & 0.008593 & 0.011360 & 0.015157 & 0.018115 & 0.000132 & 0.000185 & 0.000281 & 0.000418 \\
     130 & 0.007919 & 0.010494 & 0.014075 & 0.016789 & 0.000117 & 0.000193 & 0.000276 & 0.000412 \\
     140 & 0.007373 & 0.009770 & 0.013062 & 0.015599 & 0.000105 & 0.000143 & 0.000224 & 0.000348 \\
     150 & 0.006883 & 0.009105 & 0.012201 & 0.014561 & 0.000102 & 0.000154 & 0.000235 & 0.000368 \\
     160 & 0.006453 & 0.008527 & 0.011409 & 0.013678 & 0.000111 & 0.000142 & 0.000247 & 0.000392 \\
     170 & 0.006071 & 0.008048 & 0.010753 & 0.012889 & 0.000077 & 0.000126 & 0.000203 & 0.000263 \\
     180 & 0.005741 & 0.007606 & 0.010169 & 0.012198 & 0.000078 & 0.000124 & 0.000215 & 0.000298 \\
     190 & 0.005426 & 0.007185 & 0.009627 & 0.011511 & 0.000084 & 0.000122 & 0.000207 & 0.000280 \\
     200 & 0.005169 & 0.006854 & 0.009178 & 0.010955 & 0.000071 & 0.000102 & 0.000167 & 0.000262 \\
     210 & 0.004942 & 0.006542 & 0.008748 & 0.010466 & 0.000067 & 0.000112 & 0.000178 & 0.000253 \\
     220 & 0.004710 & 0.006233 & 0.008337 & 0.010005 & 0.000063 & 0.000091 & 0.000161 & 0.000242 \\
     230 & 0.004500 & 0.005972 & 0.008006 & 0.009560 & 0.000064 & 0.000099 & 0.000162 & 0.000220 \\
     240 & 0.004309 & 0.005715 & 0.007667 & 0.009196 & 0.000063 & 0.000096 & 0.000151 & 0.000233 \\
     250 & 0.004148 & 0.005500 & 0.007370 & 0.008818 & 0.000059 & 0.000086 & 0.000150 & 0.000186 \\
     260 & 0.003986 & 0.005283 & 0.007073 & 0.008479 & 0.000049 & 0.000076 & 0.000137 & 0.000199 \\
     270 & 0.003834 & 0.005083 & 0.006809 & 0.008153 & 0.000048 & 0.000074 & 0.000130 & 0.000185 \\
     280 & 0.003702 & 0.004906 & 0.006573 & 0.007869 & 0.000049 & 0.000074 & 0.000118 & 0.000194 \\
     290 & 0.003577 & 0.004748 & 0.006358 & 0.007602 & 0.000053 & 0.000081 & 0.000127 & 0.000171 \\
     300 & 0.003457 & 0.004583 & 0.006140 & 0.007360 & 0.000049 & 0.000072 & 0.000126 & 0.000180 \\
     310 & 0.003345 & 0.004435 & 0.005942 & 0.007107 & 0.000044 & 0.000064 & 0.000117 & 0.000174 \\
     320 & 0.003241 & 0.004302 & 0.005772 & 0.006915 & 0.000051 & 0.000067 & 0.000116 & 0.000164 \\
     330 & 0.003145 & 0.004173 & 0.005590 & 0.006682 & 0.000039 & 0.000067 & 0.000119 & 0.000156 \\
     340 & 0.003051 & 0.004047 & 0.005413 & 0.006486 & 0.000044 & 0.000065 & 0.000109 & 0.000153 \\
     350 & 0.002964 & 0.003929 & 0.005261 & 0.006327 & 0.000038 & 0.000064 & 0.000106 & 0.000153 \\
     360 & 0.002887 & 0.003826 & 0.005119 & 0.006132 & 0.000040 & 0.000065 & 0.000097 & 0.000126 \\
     370 & 0.002808 & 0.003725 & 0.005005 & 0.005994 & 0.000037 & 0.000054 & 0.000096 & 0.000139 \\
     380 & 0.002736 & 0.003627 & 0.004863 & 0.005830 & 0.000037 & 0.000060 & 0.000097 & 0.000144 \\
     390 & 0.002663 & 0.003537 & 0.004752 & 0.005690 & 0.000037 & 0.000056 & 0.000090 & 0.000135 \\
     400 & 0.002601 & 0.003447 & 0.004639 & 0.005532 & 0.000037 & 0.000062 & 0.000094 & 0.000124 \\
     410 & 0.002527 & 0.003357 & 0.004514 & 0.005406 & 0.000033 & 0.000051 & 0.000095 & 0.000123 \\
     420 & 0.002467 & 0.003278 & 0.004406 & 0.005275 & 0.000038 & 0.000055 & 0.000078 & 0.000128 \\
     430 & 0.002411 & 0.003199 & 0.004288 & 0.005136 & 0.000037 & 0.000051 & 0.000095 & 0.000112 \\
     440 & 0.002358 & 0.003129 & 0.004184 & 0.005017 & 0.000033 & 0.000052 & 0.000080 & 0.000115 \\
     450 & 0.002308 & 0.003067 & 0.004123 & 0.004934 & 0.000036 & 0.000051 & 0.000090 & 0.000115 \\
\hline
\end{tabular}

\newpage
\textbf{Table 1.} Continued.
\vskip 0.5cm

\begin{tabular}{|r|r|r|r|r|r|r|r|r|}
\hline
     460 & 0.002258 & 0.002997 & 0.004027 & 0.004833 & 0.000028 & 0.000049 & 0.000081 & 0.000111 \\
     470 & 0.002210 & 0.002933 & 0.003937 & 0.004725 & 0.000031 & 0.000050 & 0.000079 & 0.000126 \\
     480 & 0.002166 & 0.002875 & 0.003854 & 0.004624 & 0.000030 & 0.000050 & 0.000075 & 0.000115 \\
     490 & 0.002122 & 0.002810 & 0.003771 & 0.004528 & 0.000030 & 0.000041 & 0.000070 & 0.000097 \\
     500 & 0.002078 & 0.002763 & 0.003699 & 0.004432 & 0.000032 & 0.000043 & 0.000076 & 0.000122 \\
     510 & 0.002038 & 0.002710 & 0.003645 & 0.004374 & 0.000029 & 0.000043 & 0.000074 & 0.000107 \\
     520 & 0.001999 & 0.002652 & 0.003554 & 0.004268 & 0.000028 & 0.000045 & 0.000072 & 0.000112 \\
     530 & 0.001966 & 0.002605 & 0.003487 & 0.004178 & 0.000026 & 0.000039 & 0.000061 & 0.000093 \\
     540 & 0.001927 & 0.002561 & 0.003430 & 0.004108 & 0.000029 & 0.000043 & 0.000071 & 0.000104 \\
     550 & 0.001890 & 0.002506 & 0.003367 & 0.004041 & 0.000027 & 0.000041 & 0.000070 & 0.000099 \\
     560 & 0.001857 & 0.002466 & 0.003310 & 0.003980 & 0.000028 & 0.000046 & 0.000067 & 0.000098 \\
     570 & 0.001819 & 0.002414 & 0.003229 & 0.003875 & 0.000027 & 0.000039 & 0.000064 & 0.000093 \\
     580 & 0.001791 & 0.002382 & 0.003199 & 0.003829 & 0.000027 & 0.000034 & 0.000067 & 0.000092 \\
     590 & 0.001766 & 0.002344 & 0.003142 & 0.003776 & 0.000025 & 0.000039 & 0.000055 & 0.000096 \\
     600 & 0.001733 & 0.002297 & 0.003087 & 0.003701 & 0.000023 & 0.000040 & 0.000064 & 0.000093 \\
     610 & 0.001705 & 0.002264 & 0.003035 & 0.003642 & 0.000023 & 0.000036 & 0.000058 & 0.000089 \\
     620 & 0.001677 & 0.002227 & 0.002978 & 0.003571 & 0.000022 & 0.000035 & 0.000057 & 0.000086 \\
     630 & 0.001649 & 0.002193 & 0.002936 & 0.003520 & 0.000022 & 0.000034 & 0.000064 & 0.000084 \\
     640 & 0.001620 & 0.002151 & 0.002892 & 0.003471 & 0.000024 & 0.000033 & 0.000058 & 0.000093 \\
     650 & 0.001601 & 0.002121 & 0.002851 & 0.003415 & 0.000022 & 0.000038 & 0.000059 & 0.000094 \\
     660 & 0.001571 & 0.002089 & 0.002805 & 0.003366 & 0.000022 & 0.000034 & 0.000053 & 0.000083 \\
     670 & 0.001553 & 0.002063 & 0.002765 & 0.003318 & 0.000021 & 0.000034 & 0.000051 & 0.000081 \\
     680 & 0.001529 & 0.002032 & 0.002726 & 0.003262 & 0.000019 & 0.000032 & 0.000056 & 0.000084 \\
     690 & 0.001506 & 0.002003 & 0.002676 & 0.003205 & 0.000021 & 0.000034 & 0.000052 & 0.000072 \\
     700 & 0.001487 & 0.001976 & 0.002653 & 0.003180 & 0.000020 & 0.000029 & 0.000054 & 0.000080 \\
     710 & 0.001465 & 0.001944 & 0.002612 & 0.003137 & 0.000021 & 0.000030 & 0.000050 & 0.000079 \\
     720 & 0.001446 & 0.001919 & 0.002581 & 0.003088 & 0.000021 & 0.000030 & 0.000048 & 0.000068 \\
     730 & 0.001427 & 0.001890 & 0.002536 & 0.003044 & 0.000020 & 0.000027 & 0.000044 & 0.000069 \\
     740 & 0.001407 & 0.001870 & 0.002509 & 0.003005 & 0.000018 & 0.000030 & 0.000046 & 0.000070 \\
     750 & 0.001384 & 0.001838 & 0.002461 & 0.002958 & 0.000020 & 0.000028 & 0.000049 & 0.000072 \\
     760 & 0.001365 & 0.001815 & 0.002440 & 0.002926 & 0.000018 & 0.000028 & 0.000050 & 0.000069 \\
     770 & 0.001353 & 0.001795 & 0.002408 & 0.002886 & 0.000020 & 0.000029 & 0.000049 & 0.000068 \\
     780 & 0.001330 & 0.001769 & 0.002376 & 0.002847 & 0.000018 & 0.000028 & 0.000046 & 0.000075 \\
     790 & 0.001317 & 0.001748 & 0.002349 & 0.002815 & 0.000018 & 0.000029 & 0.000045 & 0.000065 \\
     800 & 0.001301 & 0.001727 & 0.002316 & 0.002770 & 0.000017 & 0.000028 & 0.000049 & 0.000072 \\
     810 & 0.001286 & 0.001709 & 0.002297 & 0.002752 & 0.000016 & 0.000023 & 0.000039 & 0.000055 \\
     820 & 0.001270 & 0.001685 & 0.002264 & 0.002718 & 0.000016 & 0.000024 & 0.000043 & 0.000071 \\
     830 & 0.001252 & 0.001661 & 0.002229 & 0.002674 & 0.000017 & 0.000028 & 0.000050 & 0.000064 \\
     840 & 0.001241 & 0.001649 & 0.002206 & 0.002651 & 0.000017 & 0.000026 & 0.000041 & 0.000064 \\
     850 & 0.001225 & 0.001627 & 0.002184 & 0.002621 & 0.000017 & 0.000026 & 0.000043 & 0.000062 \\
     860 & 0.001210 & 0.001606 & 0.002152 & 0.002581 & 0.000019 & 0.000028 & 0.000043 & 0.000064 \\
     870 & 0.001196 & 0.001583 & 0.002128 & 0.002547 & 0.000016 & 0.000022 & 0.000041 & 0.000062 \\
     880 & 0.001182 & 0.001568 & 0.002105 & 0.002521 & 0.000015 & 0.000024 & 0.000040 & 0.000058 \\
     890 & 0.001170 & 0.001552 & 0.002087 & 0.002500 & 0.000016 & 0.000027 & 0.000049 & 0.000064 \\
     900 & 0.001155 & 0.001536 & 0.002067 & 0.002481 & 0.000015 & 0.000024 & 0.000039 & 0.000060 \\
\hline
\end{tabular}

\newpage
\textbf{Table 1.} Continued.
\vskip 0.5cm

\begin{tabular}{|r|r|r|r|r|r|r|r|r|}
\hline
     910 & 0.001143 & 0.001519 & 0.002041 & 0.002441 & 0.000014 & 0.000026 & 0.000043 & 0.000064 \\
     920 & 0.001132 & 0.001504 & 0.002018 & 0.002417 & 0.000017 & 0.000025 & 0.000043 & 0.000056 \\
     930 & 0.001119 & 0.001488 & 0.001996 & 0.002393 & 0.000014 & 0.000022 & 0.000041 & 0.000068 \\
     940 & 0.001109 & 0.001473 & 0.001972 & 0.002361 & 0.000015 & 0.000021 & 0.000038 & 0.000054 \\
     950 & 0.001093 & 0.001453 & 0.001952 & 0.002342 & 0.000015 & 0.000023 & 0.000042 & 0.000061 \\
     960 & 0.001080 & 0.001434 & 0.001928 & 0.002314 & 0.000015 & 0.000024 & 0.000040 & 0.000052 \\
     970 & 0.001071 & 0.001420 & 0.001907 & 0.002288 & 0.000017 & 0.000024 & 0.000041 & 0.000057 \\
     980 & 0.001059 & 0.001406 & 0.001895 & 0.002277 & 0.000014 & 0.000022 & 0.000040 & 0.000053 \\
     990 & 0.001050 & 0.001395 & 0.001870 & 0.002244 & 0.000014 & 0.000022 & 0.000033 & 0.000052 \\
    1000 & 0.001041 & 0.001384 & 0.001861 & 0.002228 & 0.000015 & 0.000025 & 0.000042 & 0.000055 \\
\hline
   10000 & 0.000104 & 0.000138 & 0.000186 & 0.000223 & 0.000001 & 0.000002 & 0.000004 & 0.000006 \\
\hline
\end{tabular}

\newpage

\begin{table}[!ht]
\begin{center}
\caption{Simulations for the normal law $N(0,1)$.
         $n$: sample size.\\
         $\bar{K}_{0.90}$, $\bar{K}_{0.95}$, $\bar{K}_{0.98}$, $\bar{K}_{0.99}$: mean values of estimated quantiles over 100 experiments.\\
         $S_{K_{0.90}}$, $S_{K_{0.95}}$, $S_{K_{0.98}}$, $S_{K_{0.99}}$: estimated standard deviations of the quantiles distributions.}
\label{TAB2}
\end{center}
\end{table}
\vspace*{-0.75cm}
\begin{tabular}{|r|r|r|r|r|r|r|r|r|}
\hline
 $n$ & $\bar{K}_{0.90}$ & $\bar{K}_{0.95}$ & $\bar{K}_{0.98}$ & $\bar{K}_{0.99}$
     & $S_{K_{0.90}}$ & $S_{K_{0.95}}$ & $S_{K_{0.98}}$ & $S_{K_{0.99}}$ \\
\hline
       3 & 0.206148 & 0.227731 & 0.240969 & 0.245439 & 0.001423 & 0.001022 & 0.000689 & 0.000456 \\
       4 & 0.176247 & 0.220096 & 0.260605 & 0.281725 & 0.002109 & 0.002444 & 0.002400 & 0.002368 \\
       5 & 0.153131 & 0.193065 & 0.238453 & 0.266296 & 0.001812 & 0.002268 & 0.003269 & 0.003323 \\
       6 & 0.136157 & 0.173063 & 0.215660 & 0.243214 & 0.001734 & 0.002212 & 0.003252 & 0.003874 \\
       7 & 0.123281 & 0.157455 & 0.197204 & 0.223840 & 0.001479 & 0.002202 & 0.002835 & 0.003373 \\
       8 & 0.111374 & 0.142651 & 0.180713 & 0.206476 & 0.001447 & 0.001952 & 0.003128 & 0.003745 \\
       9 & 0.102793 & 0.131712 & 0.167606 & 0.192213 & 0.001246 & 0.001594 & 0.002683 & 0.003199 \\
      10 & 0.094767 & 0.121944 & 0.155632 & 0.179397 & 0.001168 & 0.001666 & 0.002249 & 0.003116 \\
      11 & 0.088537 & 0.113730 & 0.145479 & 0.167983 & 0.001080 & 0.001417 & 0.002225 & 0.003192 \\
      12 & 0.082536 & 0.106016 & 0.136342 & 0.157983 & 0.001127 & 0.001471 & 0.002004 & 0.002733 \\
      13 & 0.077855 & 0.100089 & 0.128930 & 0.149682 & 0.000976 & 0.001377 & 0.002225 & 0.003116 \\
      14 & 0.073353 & 0.094355 & 0.121794 & 0.141797 & 0.001000 & 0.001486 & 0.002356 & 0.003088 \\
      15 & 0.069663 & 0.089547 & 0.115307 & 0.134313 & 0.000766 & 0.001277 & 0.001811 & 0.002797 \\
      16 & 0.065954 & 0.084925 & 0.109532 & 0.127891 & 0.000838 & 0.001189 & 0.001799 & 0.002688 \\
      17 & 0.063034 & 0.080967 & 0.104286 & 0.121876 & 0.000829 & 0.001105 & 0.001908 & 0.002463 \\
      18 & 0.060037 & 0.077133 & 0.100098 & 0.116555 & 0.000719 & 0.001091 & 0.001589 & 0.002402 \\
      19 & 0.057690 & 0.074178 & 0.095784 & 0.111988 & 0.000669 & 0.000995 & 0.001640 & 0.002282 \\
      20 & 0.055160 & 0.070931 & 0.091782 & 0.107161 & 0.000602 & 0.001065 & 0.001546 & 0.002367 \\
      21 & 0.053234 & 0.068194 & 0.088279 & 0.103222 & 0.000589 & 0.000972 & 0.001720 & 0.002399 \\
      22 & 0.051111 & 0.065554 & 0.084898 & 0.099353 & 0.000566 & 0.000842 & 0.001475 & 0.001799 \\
      23 & 0.049378 & 0.063333 & 0.082050 & 0.096044 & 0.000602 & 0.000925 & 0.001492 & 0.001675 \\
      24 & 0.047672 & 0.061163 & 0.079168 & 0.092479 & 0.000680 & 0.000976 & 0.001557 & 0.002070 \\
      25 & 0.046052 & 0.059073 & 0.076322 & 0.089272 & 0.000582 & 0.000838 & 0.001330 & 0.001855 \\
      26 & 0.044557 & 0.057052 & 0.073518 & 0.086034 & 0.000567 & 0.000768 & 0.001238 & 0.001484 \\
      27 & 0.043118 & 0.055299 & 0.071604 & 0.083819 & 0.000537 & 0.000781 & 0.001316 & 0.001738 \\
      28 & 0.041852 & 0.053532 & 0.069190 & 0.080993 & 0.000475 & 0.000709 & 0.001165 & 0.001732 \\
      29 & 0.040710 & 0.052073 & 0.067286 & 0.078917 & 0.000479 & 0.000726 & 0.001115 & 0.001827 \\
      30 & 0.039579 & 0.050613 & 0.065551 & 0.076793 & 0.000470 & 0.000632 & 0.001084 & 0.001740 \\
      31 & 0.038442 & 0.049161 & 0.063557 & 0.074347 & 0.000448 & 0.000621 & 0.001051 & 0.001446 \\
      32 & 0.037411 & 0.047811 & 0.061803 & 0.072384 & 0.000410 & 0.000621 & 0.001012 & 0.001540 \\
      33 & 0.036456 & 0.046615 & 0.060265 & 0.070494 & 0.000464 & 0.000623 & 0.001030 & 0.001400 \\
      34 & 0.035545 & 0.045467 & 0.058717 & 0.068982 & 0.000421 & 0.000602 & 0.000886 & 0.001413 \\
      35 & 0.034782 & 0.044340 & 0.057245 & 0.066953 & 0.000454 & 0.000595 & 0.001068 & 0.001458 \\
      36 & 0.033925 & 0.043316 & 0.055951 & 0.065549 & 0.000395 & 0.000564 & 0.000877 & 0.001153 \\
      37 & 0.033158 & 0.042346 & 0.054555 & 0.063907 & 0.000400 & 0.000642 & 0.000892 & 0.001175 \\
      38 & 0.032366 & 0.041311 & 0.053444 & 0.062713 & 0.000397 & 0.000632 & 0.000945 & 0.001404 \\
      39 & 0.031691 & 0.040411 & 0.052273 & 0.061247 & 0.000403 & 0.000630 & 0.000988 & 0.001345 \\
      40 & 0.030907 & 0.039486 & 0.051016 & 0.059674 & 0.000349 & 0.000507 & 0.000737 & 0.001080 \\
      41 & 0.030470 & 0.038804 & 0.050175 & 0.058859 & 0.000375 & 0.000551 & 0.000989 & 0.001338 \\
      42 & 0.029769 & 0.037979 & 0.048965 & 0.057236 & 0.000335 & 0.000551 & 0.000792 & 0.001158 \\
      43 & 0.029212 & 0.037180 & 0.047996 & 0.056155 & 0.000357 & 0.000489 & 0.000840 & 0.001196 \\
      44 & 0.028562 & 0.036398 & 0.047058 & 0.055077 & 0.000337 & 0.000538 & 0.000892 & 0.001235 \\
      45 & 0.028021 & 0.035723 & 0.046013 & 0.053733 & 0.000287 & 0.000457 & 0.000768 & 0.001158 \\
\hline
\end{tabular}

\newpage
\textbf{Table 2.} Continued.
\vskip 0.5cm

\begin{tabular}{|r|r|r|r|r|r|r|r|r|}
\hline
      46 & 0.027504 & 0.034977 & 0.045128 & 0.052973 & 0.000332 & 0.000496 & 0.000797 & 0.001238 \\
      47 & 0.026981 & 0.034399 & 0.044468 & 0.052066 & 0.000317 & 0.000456 & 0.000664 & 0.000909 \\
      48 & 0.026539 & 0.033824 & 0.043593 & 0.051103 & 0.000304 & 0.000475 & 0.000769 & 0.001128 \\
      49 & 0.026072 & 0.033144 & 0.042658 & 0.050046 & 0.000283 & 0.000427 & 0.000674 & 0.000952 \\
      50 & 0.025603 & 0.032559 & 0.041924 & 0.049126 & 0.000284 & 0.000385 & 0.000723 & 0.001093 \\
      51 & 0.025170 & 0.032012 & 0.041354 & 0.048368 & 0.000294 & 0.000463 & 0.000723 & 0.001073 \\
      52 & 0.024719 & 0.031493 & 0.040598 & 0.047669 & 0.000271 & 0.000391 & 0.000664 & 0.001032 \\
      53 & 0.024357 & 0.030968 & 0.039957 & 0.046826 & 0.000276 & 0.000425 & 0.000722 & 0.001012 \\
      54 & 0.023921 & 0.030456 & 0.039183 & 0.045895 & 0.000258 & 0.000393 & 0.000729 & 0.001036 \\
      55 & 0.023566 & 0.029956 & 0.038575 & 0.045169 & 0.000286 & 0.000427 & 0.000683 & 0.000984 \\
      56 & 0.023173 & 0.029461 & 0.038006 & 0.044487 & 0.000237 & 0.000376 & 0.000699 & 0.000900 \\
      57 & 0.022878 & 0.029069 & 0.037395 & 0.043870 & 0.000259 & 0.000385 & 0.000655 & 0.000927 \\
      58 & 0.022505 & 0.028573 & 0.036883 & 0.043208 & 0.000294 & 0.000424 & 0.000634 & 0.000909 \\
      59 & 0.022182 & 0.028209 & 0.036341 & 0.042576 & 0.000242 & 0.000373 & 0.000608 & 0.000847 \\
      60 & 0.021792 & 0.027644 & 0.035612 & 0.041752 & 0.000290 & 0.000375 & 0.000614 & 0.000861 \\
      61 & 0.021529 & 0.027330 & 0.035223 & 0.041278 & 0.000225 & 0.000363 & 0.000579 & 0.000775 \\
      62 & 0.021162 & 0.026857 & 0.034567 & 0.040376 & 0.000245 & 0.000377 & 0.000591 & 0.000847 \\
      63 & 0.020945 & 0.026563 & 0.034225 & 0.040067 & 0.000230 & 0.000340 & 0.000558 & 0.000868 \\
      64 & 0.020709 & 0.026250 & 0.033690 & 0.039466 & 0.000237 & 0.000397 & 0.000602 & 0.000782 \\
      65 & 0.020368 & 0.025842 & 0.033178 & 0.038883 & 0.000256 & 0.000344 & 0.000515 & 0.000722 \\
      66 & 0.020081 & 0.025505 & 0.032858 & 0.038502 & 0.000235 & 0.000342 & 0.000491 & 0.000801 \\
      67 & 0.019822 & 0.025140 & 0.032367 & 0.037828 & 0.000236 & 0.000350 & 0.000497 & 0.000671 \\
      68 & 0.019617 & 0.024870 & 0.032023 & 0.037465 & 0.000249 & 0.000334 & 0.000581 & 0.000828 \\
      69 & 0.019337 & 0.024498 & 0.031553 & 0.037002 & 0.000241 & 0.000342 & 0.000628 & 0.000860 \\
      70 & 0.019047 & 0.024098 & 0.030976 & 0.036225 & 0.000202 & 0.000316 & 0.000484 & 0.000656 \\
      71 & 0.018846 & 0.023871 & 0.030774 & 0.035986 & 0.000202 & 0.000339 & 0.000499 & 0.000801 \\
      72 & 0.018614 & 0.023588 & 0.030298 & 0.035473 & 0.000195 & 0.000276 & 0.000474 & 0.000691 \\
      73 & 0.018388 & 0.023307 & 0.029923 & 0.035058 & 0.000160 & 0.000289 & 0.000468 & 0.000762 \\
      74 & 0.018165 & 0.022995 & 0.029629 & 0.034691 & 0.000203 & 0.000322 & 0.000469 & 0.000785 \\
      75 & 0.017973 & 0.022792 & 0.029273 & 0.034279 & 0.000209 & 0.000302 & 0.000517 & 0.000867 \\
      76 & 0.017768 & 0.022478 & 0.028907 & 0.033877 & 0.000214 & 0.000307 & 0.000491 & 0.000662 \\
      77 & 0.017562 & 0.022266 & 0.028636 & 0.033480 & 0.000190 & 0.000294 & 0.000464 & 0.000601 \\
      78 & 0.017347 & 0.021952 & 0.028287 & 0.033076 & 0.000206 & 0.000304 & 0.000483 & 0.000657 \\
      79 & 0.017169 & 0.021757 & 0.027986 & 0.032782 & 0.000200 & 0.000306 & 0.000506 & 0.000681 \\
      80 & 0.016962 & 0.021429 & 0.027523 & 0.032168 & 0.000194 & 0.000301 & 0.000481 & 0.000632 \\
      81 & 0.016794 & 0.021259 & 0.027344 & 0.032046 & 0.000193 & 0.000333 & 0.000521 & 0.000711 \\
      82 & 0.016587 & 0.020979 & 0.026945 & 0.031550 & 0.000170 & 0.000271 & 0.000465 & 0.000628 \\
      83 & 0.016428 & 0.020782 & 0.026712 & 0.031312 & 0.000187 & 0.000260 & 0.000449 & 0.000712 \\
      84 & 0.016240 & 0.020571 & 0.026359 & 0.030807 & 0.000180 & 0.000275 & 0.000483 & 0.000707 \\
      85 & 0.016076 & 0.020330 & 0.026066 & 0.030547 & 0.000174 & 0.000291 & 0.000514 & 0.000823 \\
      86 & 0.015922 & 0.020125 & 0.025828 & 0.030202 & 0.000187 & 0.000278 & 0.000435 & 0.000642 \\
      87 & 0.015723 & 0.019908 & 0.025545 & 0.029892 & 0.000181 & 0.000286 & 0.000469 & 0.000645 \\
      88 & 0.015587 & 0.019710 & 0.025353 & 0.029600 & 0.000201 & 0.000265 & 0.000469 & 0.000680 \\
      89 & 0.015432 & 0.019527 & 0.025107 & 0.029337 & 0.000178 & 0.000264 & 0.000394 & 0.000573 \\
      90 & 0.015265 & 0.019309 & 0.024773 & 0.029009 & 0.000182 & 0.000267 & 0.000409 & 0.000632 \\
\hline
\end{tabular}

\newpage
\textbf{Table 2.} Continued.
\vskip 0.5cm

\begin{tabular}{|r|r|r|r|r|r|r|r|r|}
\hline
      91 & 0.015112 & 0.019104 & 0.024485 & 0.028674 & 0.000181 & 0.000262 & 0.000414 & 0.000590 \\
      92 & 0.014991 & 0.018947 & 0.024253 & 0.028369 & 0.000175 & 0.000261 & 0.000445 & 0.000672 \\
      93 & 0.014830 & 0.018752 & 0.023998 & 0.028147 & 0.000174 & 0.000260 & 0.000465 & 0.000606 \\
      94 & 0.014673 & 0.018556 & 0.023807 & 0.027843 & 0.000158 & 0.000244 & 0.000417 & 0.000627 \\
      95 & 0.014543 & 0.018352 & 0.023520 & 0.027501 & 0.000166 & 0.000251 & 0.000402 & 0.000581 \\
      96 & 0.014418 & 0.018171 & 0.023371 & 0.027314 & 0.000152 & 0.000232 & 0.000405 & 0.000613 \\
      97 & 0.014249 & 0.018026 & 0.023127 & 0.027003 & 0.000159 & 0.000234 & 0.000426 & 0.000662 \\
      98 & 0.014145 & 0.017878 & 0.022875 & 0.026742 & 0.000177 & 0.000241 & 0.000414 & 0.000590 \\
      99 & 0.014033 & 0.017695 & 0.022721 & 0.026638 & 0.000149 & 0.000240 & 0.000372 & 0.000532 \\
     100 & 0.013912 & 0.017552 & 0.022499 & 0.026394 & 0.000165 & 0.000248 & 0.000405 & 0.000634 \\
\hline
     110 & 0.012782 & 0.016139 & 0.020678 & 0.024151 & 0.000152 & 0.000215 & 0.000344 & 0.000458 \\
     120 & 0.011799 & 0.014855 & 0.019032 & 0.022210 & 0.000115 & 0.000196 & 0.000331 & 0.000437 \\
     130 & 0.010972 & 0.013802 & 0.017673 & 0.020649 & 0.000127 & 0.000186 & 0.000275 & 0.000473 \\
     140 & 0.010277 & 0.012932 & 0.016503 & 0.019234 & 0.000113 & 0.000192 & 0.000281 & 0.000394 \\
     150 & 0.009661 & 0.012134 & 0.015499 & 0.018072 & 0.000103 & 0.000155 & 0.000266 & 0.000393 \\
     160 & 0.009094 & 0.011421 & 0.014616 & 0.017020 & 0.000105 & 0.000158 & 0.000239 & 0.000381 \\
     170 & 0.008625 & 0.010829 & 0.013816 & 0.016068 & 0.000085 & 0.000134 & 0.000252 & 0.000366 \\
     180 & 0.008171 & 0.010249 & 0.013074 & 0.015266 & 0.000093 & 0.000146 & 0.000226 & 0.000306 \\
     190 & 0.007800 & 0.009781 & 0.012456 & 0.014526 & 0.000086 & 0.000130 & 0.000207 & 0.000306 \\
     200 & 0.007437 & 0.009308 & 0.011870 & 0.013828 & 0.000076 & 0.000116 & 0.000194 & 0.000287 \\
     210 & 0.007101 & 0.008911 & 0.011343 & 0.013223 & 0.000081 & 0.000115 & 0.000186 & 0.000258 \\
     220 & 0.006806 & 0.008513 & 0.010829 & 0.012609 & 0.000074 & 0.000118 & 0.000192 & 0.000279 \\
     230 & 0.006544 & 0.008183 & 0.010406 & 0.012120 & 0.000068 & 0.000097 & 0.000166 & 0.000244 \\
     240 & 0.006292 & 0.007861 & 0.009999 & 0.011630 & 0.000062 & 0.000101 & 0.000163 & 0.000232 \\
     250 & 0.006060 & 0.007573 & 0.009640 & 0.011220 & 0.000061 & 0.000105 & 0.000167 & 0.000236 \\
     260 & 0.005849 & 0.007294 & 0.009282 & 0.010826 & 0.000069 & 0.000094 & 0.000158 & 0.000219 \\
     270 & 0.005637 & 0.007041 & 0.008924 & 0.010367 & 0.000064 & 0.000102 & 0.000148 & 0.000191 \\
     280 & 0.005463 & 0.006825 & 0.008668 & 0.010077 & 0.000059 & 0.000085 & 0.000140 & 0.000210 \\
     290 & 0.005284 & 0.006594 & 0.008368 & 0.009739 & 0.000054 & 0.000089 & 0.000137 & 0.000194 \\
     300 & 0.005125 & 0.006385 & 0.008094 & 0.009432 & 0.000051 & 0.000085 & 0.000129 & 0.000171 \\
     310 & 0.004970 & 0.006199 & 0.007860 & 0.009136 & 0.000059 & 0.000082 & 0.000121 & 0.000176 \\
     320 & 0.004825 & 0.006012 & 0.007631 & 0.008856 & 0.000047 & 0.000071 & 0.000127 & 0.000199 \\
     330 & 0.004685 & 0.005832 & 0.007392 & 0.008598 & 0.000053 & 0.000076 & 0.000118 & 0.000168 \\
     340 & 0.004572 & 0.005692 & 0.007204 & 0.008396 & 0.000050 & 0.000074 & 0.000116 & 0.000159 \\
     350 & 0.004449 & 0.005532 & 0.007010 & 0.008137 & 0.000046 & 0.000071 & 0.000114 & 0.000171 \\
     360 & 0.004324 & 0.005383 & 0.006808 & 0.007920 & 0.000047 & 0.000064 & 0.000113 & 0.000157 \\
     370 & 0.004224 & 0.005259 & 0.006654 & 0.007745 & 0.000042 & 0.000067 & 0.000097 & 0.000156 \\
     380 & 0.004117 & 0.005117 & 0.006471 & 0.007535 & 0.000045 & 0.000061 & 0.000091 & 0.000131 \\
     390 & 0.004015 & 0.004990 & 0.006310 & 0.007345 & 0.000044 & 0.000064 & 0.000105 & 0.000150 \\
     400 & 0.003924 & 0.004889 & 0.006189 & 0.007190 & 0.000040 & 0.000061 & 0.000096 & 0.000148 \\
     410 & 0.003839 & 0.004768 & 0.006034 & 0.007003 & 0.000040 & 0.000057 & 0.000093 & 0.000158 \\
     420 & 0.003745 & 0.004653 & 0.005875 & 0.006841 & 0.000035 & 0.000060 & 0.000102 & 0.000136 \\
     430 & 0.003667 & 0.004559 & 0.005763 & 0.006691 & 0.000042 & 0.000056 & 0.000086 & 0.000129 \\
     440 & 0.003589 & 0.004456 & 0.005625 & 0.006535 & 0.000041 & 0.000060 & 0.000096 & 0.000132 \\
     450 & 0.003515 & 0.004373 & 0.005522 & 0.006427 & 0.000037 & 0.000056 & 0.000097 & 0.000129 \\
\hline
\end{tabular}

\newpage
\textbf{Table 2.} Continued.
\vskip 0.5cm

\begin{tabular}{|r|r|r|r|r|r|r|r|r|}
\hline
     460 & 0.003445 & 0.004278 & 0.005416 & 0.006302 & 0.000036 & 0.000054 & 0.000097 & 0.000129 \\
     470 & 0.003379 & 0.004189 & 0.005281 & 0.006133 & 0.000038 & 0.000056 & 0.000081 & 0.000105 \\
     480 & 0.003318 & 0.004113 & 0.005189 & 0.006031 & 0.000033 & 0.000048 & 0.000087 & 0.000112 \\
     490 & 0.003250 & 0.004033 & 0.005100 & 0.005910 & 0.000033 & 0.000048 & 0.000075 & 0.000111 \\
     500 & 0.003186 & 0.003945 & 0.004973 & 0.005791 & 0.000032 & 0.000053 & 0.000077 & 0.000119 \\
     510 & 0.003132 & 0.003885 & 0.004904 & 0.005699 & 0.000030 & 0.000053 & 0.000082 & 0.000120 \\
     520 & 0.003073 & 0.003812 & 0.004811 & 0.005572 & 0.000039 & 0.000054 & 0.000077 & 0.000123 \\
     530 & 0.003018 & 0.003737 & 0.004717 & 0.005480 & 0.000030 & 0.000045 & 0.000075 & 0.000109 \\
     540 & 0.002965 & 0.003676 & 0.004638 & 0.005388 & 0.000033 & 0.000050 & 0.000071 & 0.000101 \\
     550 & 0.002917 & 0.003615 & 0.004550 & 0.005297 & 0.000030 & 0.000043 & 0.000067 & 0.000098 \\
     560 & 0.002862 & 0.003550 & 0.004478 & 0.005192 & 0.000028 & 0.000049 & 0.000068 & 0.000098 \\
     570 & 0.002818 & 0.003488 & 0.004407 & 0.005125 & 0.000030 & 0.000045 & 0.000071 & 0.000110 \\
     580 & 0.002778 & 0.003440 & 0.004334 & 0.005025 & 0.000027 & 0.000040 & 0.000073 & 0.000100 \\
     590 & 0.002728 & 0.003378 & 0.004254 & 0.004932 & 0.000028 & 0.000043 & 0.000069 & 0.000099 \\
     600 & 0.002686 & 0.003326 & 0.004195 & 0.004869 & 0.000025 & 0.000043 & 0.000066 & 0.000095 \\
     610 & 0.002649 & 0.003280 & 0.004128 & 0.004790 & 0.000024 & 0.000039 & 0.000060 & 0.000091 \\
     620 & 0.002611 & 0.003231 & 0.004059 & 0.004722 & 0.000029 & 0.000042 & 0.000061 & 0.000091 \\
     630 & 0.002568 & 0.003175 & 0.003993 & 0.004633 & 0.000023 & 0.000041 & 0.000060 & 0.000087 \\
     640 & 0.002530 & 0.003126 & 0.003937 & 0.004564 & 0.000027 & 0.000041 & 0.000066 & 0.000095 \\
     650 & 0.002494 & 0.003087 & 0.003893 & 0.004519 & 0.000027 & 0.000041 & 0.000064 & 0.000085 \\
     660 & 0.002457 & 0.003038 & 0.003828 & 0.004430 & 0.000024 & 0.000038 & 0.000061 & 0.000083 \\
     670 & 0.002424 & 0.002999 & 0.003773 & 0.004371 & 0.000025 & 0.000033 & 0.000054 & 0.000087 \\
     680 & 0.002394 & 0.002961 & 0.003726 & 0.004318 & 0.000026 & 0.000036 & 0.000055 & 0.000082 \\
     690 & 0.002358 & 0.002912 & 0.003665 & 0.004247 & 0.000024 & 0.000035 & 0.000055 & 0.000088 \\
     700 & 0.002330 & 0.002878 & 0.003626 & 0.004201 & 0.000023 & 0.000032 & 0.000056 & 0.000082 \\
     710 & 0.002295 & 0.002839 & 0.003565 & 0.004130 & 0.000022 & 0.000032 & 0.000049 & 0.000074 \\
     720 & 0.002268 & 0.002804 & 0.003528 & 0.004087 & 0.000023 & 0.000034 & 0.000051 & 0.000081 \\
     730 & 0.002233 & 0.002758 & 0.003479 & 0.004033 & 0.000023 & 0.000033 & 0.000057 & 0.000083 \\
     740 & 0.002211 & 0.002734 & 0.003436 & 0.003984 & 0.000022 & 0.000034 & 0.000052 & 0.000079 \\
     750 & 0.002182 & 0.002697 & 0.003398 & 0.003936 & 0.000022 & 0.000035 & 0.000052 & 0.000088 \\
     760 & 0.002156 & 0.002661 & 0.003350 & 0.003884 & 0.000021 & 0.000028 & 0.000054 & 0.000075 \\
     770 & 0.002127 & 0.002624 & 0.003294 & 0.003811 & 0.000020 & 0.000027 & 0.000048 & 0.000072 \\
     780 & 0.002099 & 0.002596 & 0.003266 & 0.003780 & 0.000022 & 0.000031 & 0.000052 & 0.000071 \\
     790 & 0.002075 & 0.002563 & 0.003220 & 0.003730 & 0.000021 & 0.000033 & 0.000054 & 0.000077 \\
     800 & 0.002051 & 0.002535 & 0.003192 & 0.003695 & 0.000020 & 0.000032 & 0.000054 & 0.000075 \\
     810 & 0.002032 & 0.002508 & 0.003154 & 0.003655 & 0.000020 & 0.000032 & 0.000049 & 0.000072 \\
     820 & 0.002006 & 0.002476 & 0.003108 & 0.003589 & 0.000018 & 0.000028 & 0.000049 & 0.000069 \\
     830 & 0.001983 & 0.002452 & 0.003078 & 0.003567 & 0.000021 & 0.000033 & 0.000049 & 0.000078 \\
     840 & 0.001962 & 0.002418 & 0.003038 & 0.003520 & 0.000019 & 0.000028 & 0.000054 & 0.000083 \\
     850 & 0.001938 & 0.002387 & 0.003007 & 0.003472 & 0.000020 & 0.000027 & 0.000044 & 0.000057 \\
     860 & 0.001918 & 0.002367 & 0.002974 & 0.003446 & 0.000020 & 0.000026 & 0.000043 & 0.000069 \\
     870 & 0.001896 & 0.002340 & 0.002944 & 0.003412 & 0.000018 & 0.000029 & 0.000046 & 0.000069 \\
     880 & 0.001878 & 0.002316 & 0.002905 & 0.003368 & 0.000018 & 0.000027 & 0.000046 & 0.000064 \\
     890 & 0.001855 & 0.002288 & 0.002873 & 0.003327 & 0.000018 & 0.000026 & 0.000048 & 0.000064 \\
     900 & 0.001835 & 0.002263 & 0.002841 & 0.003298 & 0.000021 & 0.000028 & 0.000049 & 0.000067 \\
\hline
\end{tabular}

\newpage
\textbf{Table 2.} Continued.
\vskip 0.5cm

\begin{tabular}{|r|r|r|r|r|r|r|r|r|}
\hline
     910 & 0.001817 & 0.002241 & 0.002813 & 0.003257 & 0.000018 & 0.000027 & 0.000045 & 0.000069 \\
     920 & 0.001801 & 0.002220 & 0.002781 & 0.003222 & 0.000017 & 0.000027 & 0.000048 & 0.000066 \\
     930 & 0.001777 & 0.002195 & 0.002757 & 0.003196 & 0.000019 & 0.000026 & 0.000043 & 0.000059 \\
     940 & 0.001761 & 0.002174 & 0.002732 & 0.003162 & 0.000018 & 0.000026 & 0.000046 & 0.000066 \\
     950 & 0.001744 & 0.002149 & 0.002696 & 0.003115 & 0.000019 & 0.000025 & 0.000042 & 0.000062 \\
     960 & 0.001725 & 0.002128 & 0.002671 & 0.003100 & 0.000018 & 0.000027 & 0.000049 & 0.000062 \\
     970 & 0.001712 & 0.002110 & 0.002648 & 0.003072 & 0.000017 & 0.000028 & 0.000043 & 0.000062 \\
     980 & 0.001694 & 0.002089 & 0.002618 & 0.003036 & 0.000018 & 0.000025 & 0.000042 & 0.000063 \\
     990 & 0.001678 & 0.002070 & 0.002596 & 0.003001 & 0.000015 & 0.000024 & 0.000036 & 0.000057 \\
    1000 & 0.001664 & 0.002051 & 0.002572 & 0.002977 & 0.000015 & 0.000024 & 0.000050 & 0.000068 \\
\hline
   10000 & 0.000184 & 0.000224 & 0.000276 & 0.000317 & 0.000002 & 0.000003 & 0.000005 & 0.000006 \\
\hline
\end{tabular}

\newpage
\parindent 1cm

\section{Conclusions}

Tables \ref{TAB1} and \ref{TAB2} have been generated respectively for the uniform $U(0,1)$ and for the normal $N(0,1)$ parent distributions.
The sampling distribution of the chiral index statistic issued from an uniform law $U(a,b)$, $a<b$, does not depend on the interval $\lbrack a;b \rbrack$.
The sampling distribution of the chiral index statistic issued from a normal law $N(m,s)$ depends neither from the expectation $m$ nor on the standard deviation $s$.
It means that the chiral index allows to perform location free and scale free symmetry tests.

\vskip 0.5cm
It is expected that the tables will be used for symmetry testing.
Nevertheless, asymptotic expressions of the sampling distribution of the chiral index remain desirable.
The chiral index is easy to compute, even with a pocket calculator.
So, beginners in programming should be able to generate more tables when desired.
The following section of code in R \cite{RCRAN} computes the chiral index of a sample:
\begin{verbatim}
x <- scan(file="") # import observations from keyboard; type is double
chiral_index <- (1+cor(sort(x,decreasing=FALSE),sort(x,decreasing=TRUE)))/2
chiral_index # dsplay the value of the chiral index
\end{verbatim}

\renewcommand{\refname}{References}

\end{document}